\documentclass[]{aa}  
\usepackage{graphicx}
\usepackage[authoryear]{natbib}
\bibpunct{(}{)}{;}{a}{}{,}
\usepackage{txfonts}
\begin{document}
   \title{A search for counterparts to massive X-ray
   binaries using photometric catalogues\fnmsep\thanks{Based on
   observations collected at the South African Astronomical Observatory}} 

   \author{I.~Negueruela
          \inst{1,2}
          \and
          M. P. E. Schurch\inst{3}
	            }
   \offprints{I.~Negueruela}
   \institute{Departamento de F\'{\i}sica, Ingenier\'{\i}a de Sistemas y
  Teor\'{\i}a de la Se\~{n}al, Universidad de Alicante, Apdo. 99,
  E03080 Alicante, Spain\\
              \email{ignacio@dfists.ua.es}
\and
	  Department of Physics and Astronomy, The Open University,
  Walton Hall, Milton Keynes MK7 6AA, United Kingdom 
         \and
	 School of Physics and Astronomy, University of
          Southampton, Southampton SO17 1BJ,  United Kingdom} 
   \date{Received }

 
  \abstract{The X-ray and $\gamma$-ray observatory {\it INTEGRAL} {\rm has
  discovered large numbers of new hard X-ray sources, many of which
  are believed to be high mass X-ray binaries. However, for a significant
  fraction, their counterparts remain unidentified.}}
   {{\rm We} explore the use of photometric catalogues to find optical
  counterparts to high mass X-ray binaries and search for objects
  likely to be early-type stars within the error circles of several
  {\it INTEGRAL} {\rm sources}.}  
   {{\rm Candidates} were selected from 2MASS photometry by means of a
  reddening free $Q$ parameter. Sufficiently bright candidates were
  spectroscopically observed.}
   {Many of the candidates selected turned out to be moderately
  reddened late A or early F stars. Optically visible OB stars are
  very scarce even in these Galactic Plane fields.
  Our method is able to identify the counterpart to IGR~J16207$-$5129,
  confirmed by a {\it Chandra} {\rm localisation. We classify this
  object as a B0 supergiant}.
 {\rm In} the field of AX J1820.5$-$1434, we find a mid or early B-type star,
  but we cannot confirm it as the counterpart.\\
 For IGR~J16320$-$4751 we rule out the optically visible candidate as
  a {\rm possible} counterpart. \\
 For AX~J1700.2$-$4220, we do not find any suitable candidate within
  the {\it ASCA} {\rm error} circle. We classify HD~153295, a marginal
  candidate to be the counterpart, as B0.5\,IVe, and find a distance
  compatible with membership in Sco OB1. \\
 In the case of
  IGR~J17091$-$3624, the object associated with a variable radio
  source in the field is a late F star.\\
We derive a spectral type B0\,IIIe for HD~100199, previously
  identified as the
  counterpart to IGR~J11305$-$6256.}
   {The procedure used is able to correctly identify {\rm OB}
  stars and, in about 
  one third of the cases, may lead to the localisation of the correct
  counterpart. However, the majority of {\it INTEGRAL} {\rm error}
  circles do not contain any 
  suitable optically visible counterpart. Deep infrared searches are
  going to be necessary in order to locate the counterparts to these sources.}

\titlerunning{Counterparts to HMXBs}

   \maketitle
%

\section{Introduction}
High Mass X-ray Binaries (HMXBs) are X-ray sources powered
  by accretion on to a compact object of material coming from a
  companion massive star. HMXBs are objects of the highest
  astrophysical interest, as their study allows us to address a number
  of fundamental questions, from the masses of neutron stars to the
  structure of stellar winds \citep[e.g.,][]{kap04}. They can also be
  used to constrain models of stellar 
  evolution, binary evolution and the mechanisms for the formation
  of neutron stars and black holes. Moreover, because of their young age, when
  considered as a population, they can provide information on
  the properties of galaxies, such as their star formation rates
  \citep[e.g.,][]{grim03}.  

 During the last couple of years, the ESA
X-ray and $\gamma$-ray 
observatory {\it INTEGRAL} has been conducting a continuous monitoring
of the Galactic Plane, with especial attention to the Galactic Centre
and its surroundings. These observations have revealed large numbers of new
X-ray sources, many of which appear hard. In most cases, these
  hard spectra are believed to be due to high interstellar extinction
  resulting in the absorption of any soft components. A substantial
fraction of these sources are believed to be HMXBs, based on their
X-ray characteristics \citep[see, e.g.,][]{wal06}, and in many cases,
this has been proved by the 
discovery of their counterparts
\citep[e.g.,][]{fc04,reig05,mas06,neg}. 

Interestingly, many of these objects are believed to have OB
supergiants as mass donors. In some cases, the X-ray sources are
pulsed and orbital parameters typical of persistent Supergiant X-ray
Binaries (SGXBs) have been found \citep[e.g.,][]{bog,zurita}. In several other
cases, the systems have been identified as supergiant fast X-ray
transients (SFXTs), displaying short outbursts
\citep{sgue05,esa,smith,sidoli}. This is a very important result,
because for many years it was thought that the population of SGXBs
should be relatively small (because of evolutionary reasons) and that
a substantial fraction of its members were already known (because they
were persistent moderately bright X-ray sources). The discovery of
{\it many} new systems (currently, {\it INTEGRAL} has found or
identified more SGXBs than were previously known; cf. \citealt{wal06})
represents a 
challenge to binary star population synthesis models. Because of this,
the search for the counterparts to as many new {\it INTEGRAL} sources
as possible is a very urgent matter.

Here we report on our search for counterparts to a number of {\it
  INTEGRAL} sources believed to be HMXBs through the selection of
  candidate OB stars in their error circle based on their 2MASS \citep{skru06}
  colours. We also present new spectra and spectral classifications
  for some proposed counterparts.


\section{Methodology and observations}

 As is only natural, candidate HMXBs lie close to the Galactic
 Plane. {\it INTEGRAL} error circles have in most cases $2\arcmin$
 radii (some are slightly worse) and therefore contain many dozens of
 optical and infrared sources. In a few cases, the {\it INTEGRAL}
 source coincides with an {\it ASCA} source and then the error circle
 is smaller ($50\arcsec$ radius), though still relatively large. The
 number of possible counterparts is simply far too large to tackle.

The situation becomes somewhat better if we expect the counterpart to
a given source to be a high mass star. These objects are intrinsically
blue and intrinsically bright. Unless they are obscured by enormous
amounts of extinction, they are expected to be relatively bright in
the $K$ band, where the extinction is $\sim10$ times lower than in the
optical. Because of this, it is not unreasonable to expect to find the
counterpart among the stars detected in the 2MASS catalogue.

We have used 2MASS photometric data to search for objects within the
 X-ray error circles that might be intrinsically blue. Under the
 assumption of a standard reddening law, which is much more likely to
 hold in the infrared than in the UV/optical region
 \citep[e.g.,][]{inde05}, the observed infrared colours of stars
 can be projected along a known reddening law to the location
of their intrinsic
 colours. However, in the near-infrared, the colours of early-type
 stars are basically degenerate (cf.~\citealt{ducati}, where the
 $(H-K)_{0}$ colours of B-type and early A-type hardly cover a range
 of $0.1$ mag) and
 therefore typical 2MASS errors, of the order of
 $\sim0.03\,$--$\,0.05$~mag in a 
 given colour even for bright stars, do not allow an accurate
 de-reddening. A simpler approach, such as the calculation of an
 infrared equivalent to Johnson's $Q$ parameter, is likely to give
 similar results.

If we define the reddening-free quantity $Q = (J-H) - 1.70 (H-K_{\rm
  S})$, the intrinsic colours of early and late-type stars are such
  that they will lie clearly separated in the $Q/K_{\rm S}$ diagram
  \citep[e.g.,][]{cp05}. The 
  majority of stars in Galactic fields concentrate around $Q=0.4-0.5$,
  corresponding to field K and M stars, while early-type stars
  typically have $Q\simeq0$. 

For early-type supergiants, several factors can affect their intrinsic
 colours, among them, variations in the extinction law\footnote{The
 factor $E(J-H)/E(H-K)=1.70$ is obtained from the extinction law of
 \citet{rl85}. The extinction law of \citet{fitz99} gives a value
 $E(J-H)/E(H-K)\approx1.9$. As a matter of fact, both extinction laws
 use Johnson's $K$ band, 
 while 2MASS uses $K_{{\rm S}}$. Its shorter wavelength should, in
 principle, result
 in higher extinction, though empirical results by \citet{inde05}
 indicate a value $\approx1.8$. The adoption of
 $E(J-H)/E(H-K_{{\rm S}})=1.8$ would only affect our results
for the case of heavily reddened stars, resulting in more negative
 values of $Q$. In view of 
 this, we prefer to stick to the 1.7 
 value, as the location of stars in this diagram is well explored
 and the $Q$ values of OB stars are known to be well separated
 from those of later-type field stars.},
 infrared excesses, etc. Because of this, we preferred to check the
 $Q$ values of known SGXBs, mostly those detected by {\it INTEGRAL},
 with known counterparts. They have a range of $Q$ values, extending
 from $-0.05$ to $0.15$. Therefore we proceeded to search 
 the error circles of {\it INTEGRAL} sources selecting sources
 according to the following criteria: 

\begin{itemize}
\item We selected stars with $K_{\rm S}<11$ and $Q<0.2$ as possible OB
  supergiants (in practice, a value of $Q\la-0.1$ would indicate
  an infrared excess, but we did not find any star fulfilling this
  condition). A magnitude cutoff is reasonable in this case. If a 
  Galactic OB supergiant is fainter than this limit in $K_{\rm S}$,
  this would imply amounts of obscuration such that the $J$ band
  measurement would be unreliable (see typical values in
  Table~\ref{tab:final}). Moreover, the object would certainly
  be too faint in the optical to have been observed for this work.
\item We selected stars with $K_{\rm S}<12$ and $Q<0$ as possible Be
  stars. Be stars are characterised by strong infrared excesses and
  application of a dereddening procedure that assumes standard
  reddening results in an ``over-correction'' that leads to negative
  $Q$ values. Obviously, we could have found Be stars at
  magnitudes fainter than $K_{\rm S}=12$, but again those objects
  would be too reddened to have reliable $J$ magnitudes.
\item We rejected stars whose USNO colours were incompatible with a
  reddened OB star, i.e., objects that were too bright in the blue
  band compared to their $J$ magnitude and $(J-K)$ colours. As a
  reference, the standard reddening law implies $E(B-R)=3.4E(J-K)$ and
  so an OB star with $(J-K)\approx1$ should have $(B-R)\ga3$.
\end{itemize}

We tested this method on the fields of five SGXBs detected by {\it
  INTEGRAL} with known counterparts, all of which are relatively faint
  in the optical. In all cases, the counterpart was selected among the
  very few candidates complying with the criteria. Application of this
  method has led to the identification of an obscured OB supergiant in
  the field of the SFXT SAX~J1818.6$-$1703 \citep{ns06}. 
  However, the criteria do
  not guarantee the selection of an OB star. Reddened A-type and early
  F-type stars will still have $Q<0.2$ (A stars will actually be quite close
  to $Q=0$), while oxygen-rich AGB stars and carbon-rich giants can
  display $JHK_{\rm S}$ colours 
  indistinguishable from those of normal, reddened early-type stars
  \citep{bb88,cp05}. Moreover, all kinds of stars with infrared
  excess, such as T~Tauri stars, will have $Q<0$. Spectroscopic
  observations are thus required 
  to confirm the nature of the candidates.

 Observations of the candidate stars were carried out on the nights of 2006
 May 2nd to 8th using the unit spectrograph on the 1.9-m telescope at
 the South African Astronomical Observatory in
 Sutherland\fnmsep\footnote{{\tt
 http://www.saao.ac.za/facilities/instrumentation/gratingspec/}}. We
 used 
 gratings \#8 and \#9 to observe objects in the red region and grating
 \#6 to observe them in the blue. In all cases, we used a $1\farcs5$
 slit. 

 Grating \#8 (830 ln/mm) was used on the nights of May 4th and 6th. It
 covers the  6200\,--\,7400~\AA\ range with a resolution element
 (measured on arc 
 frames) of 1.5\AA\ around $\lambda$~7000\AA . 

Grating \#9 was used on the night of May 7th. It covers the
 6200\,--\,8900\AA\ region with a resolution element (measured on arc frames)
 of 3.2\AA\ around $\lambda$~7500\AA.

Grating \#6 (600 ln/mm) was used on the night of May 8th. It covers
the  3800\,--\,5600~\AA\ range with a resolution element (measured on arc
 frames) of 2.3\AA\ around $\lambda$~4500\AA. 

Two faint targets were later observed on the night of June 5th 2006
with the 2.6-m Nordic Optical Telescope (NOT), in La Palma (Spain),
equipped with the Andalucia Faint Object Spectrograph and Camera
(ALFOSC)\fnmsep\footnote{{\tt
    http://www.not.iac.es/instruments/alfosc/}}. We used grism \#4,
which provides a dispersion of 3\AA/pixel 
over the whole optical range.

All the spectra have been reduced with the {\em Starlink}
packages {\sc ccdpack} \citep{draper} and {\sc figaro}
\citep{shortridge} and analysed using {\sc figaro} and {\sc dipso}
\citep{howarth}.

\section{Results}

\subsection{IGR~J11305$-$6256}
 
The weak transient source \object{IGR~J11305$-$6256} was detected by
IBIS/ISGRI  in May 2004 \citep{pro04}. \citet{mas06} noted that the
catalogued  Be
star HD~100199 was within
the large ($5\arcmin$ radius) error circle, and later confirmed the
identification with a {\it SWIFT}/XRT detection (error circle of
radius only $6\arcsec$). Quoting a spectral type of B0\,IIIe from
\citet{gar77}, they estimate a distance of $\sim 3\:$kpc. We note
  that, with $Q=-0.12$,  HD~100199 is the only star standing out in
  the $Q$/$K$ diagram for the error circle, though there are a couple
  other marginal candidates.

%
   \begin{figure}
   \centering
   \resizebox{\columnwidth}{!}{\includegraphics[angle=-90, bb=100 50 450 780,clip]{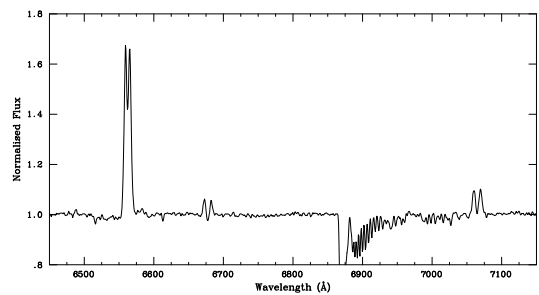}}
   \caption{Red spectrum of HD~100199, the optical
   counterpart to IGR~J11305$-$6256, showing the clearly separated
   double peaks in all emission lines. The central absorption in
   \ion{He}{i}~6678\AA\ comes below the continuum level, indicating
   that HD~100199 is a shell star.}
              \label{reddo}
    \end{figure}
%

As the only modern spectrogram of this source is the low
resolution spectrum of \citet{mas06}, we have obtained intermediate
resolution spectra of HD~100199, in order to reassess its spectral
classification. The red spectrum of HD~100199 (see Fig.~\ref{reddo})
displays double-peaked emission lines corresponding to H$\alpha$ and
\ion{He}{i}~6678, 7065\AA. \citet{mas06} report single-peaked emission
in H$\alpha$, but this is likely to be due simply to the low
resolution of their spectrum. There is a weak emission line around
$\lambda$6582\AA, which may be due to the \ion{C}{ii}~6578, 6582\AA\
doublet.

\begin{figure}
   \centering
   \resizebox{\columnwidth}{!}{\includegraphics[bb= 50 90 480 760,angle=-90]{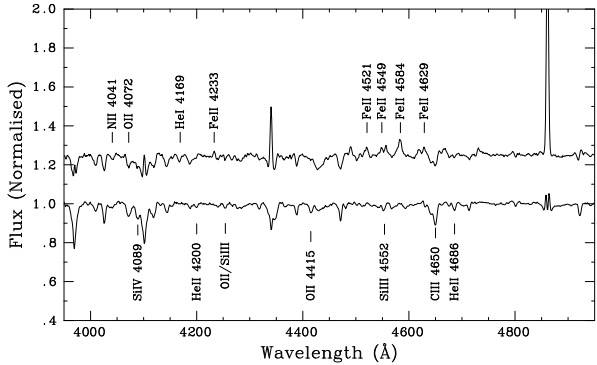}}
   \caption{Classification spectra of HD~100199 (bottom), the optical
   counterpart to IGR~J11305$-$6256, and HD~153295 (top), possible
   counterpart to  AX J1700.2$-$4220.}
              \label{bes}
    \end{figure}

The blue spectrum (see Fig.~\ref{bes}) shows double-peaked emission in
H$\beta$ and emission infilling in H$\gamma$. Four \ion{He}{ii} lines
are visible. In particular, the prominent \ion{He}{ii}~5412\AA\ line
(not shown)   
and the presence of a weak \ion{He}{ii}~4542\AA\ indicate that the
star has to be classified B0 or earlier. The fact that
\ion{He}{ii}~4542\AA$<$\ion{Si}{iii}~4552\AA\ prevents an O-type
classification and sets the spectral type at B0. The weakness of
\ion{He}{ii}~4686\AA, which is inversely correlated with luminosity
class, shows that this star is at least of moderate luminosity.
The intensity of the \ion{C}{iii}~4650\AA\ line when compared to
\ion{He}{i} lines would suggest a very high luminosity, as it is
  clearly stronger than \ion{He}{i}~4471\AA. However, this high
  luminosity is not supported by the ratio of \ion{Si}{iv} lines to
  \ion{He}{i} lines or the fact that  
\ion{He}{i}~4713$\approx$\ion{He}{ii}~4686\AA. Therefore we adopt a
luminosity class III, confirming the classification of \citet{gar77},
though noting the abnormal strength of 
\ion{C}{iii}~4650\AA. 

As the H$\alpha$ emission is not strong (EW=$-7$\AA), the contribution
of the circumstellar disk to $E(B-V)$ will likely not be very large,
and so the distance calculated by \citet{mas06} is a good
estimate. This object is likely to be a persistent low-luminosity
Be/X-ray binary.

%
%

\subsection{IGR~J16207$-$5129}

This source appears in the first {\it INTEGRAL} catalogue
\citep{bird04} as INTEGRAL1 22. \citet{tomsick} selected it as a
seemingly persistent source, 
likely to be a HMXB, and observed it with  {\it Chandra}, resulting in
an accurate localisation. 
\citet{mastel} have obtained a spectrum of the only object in the {\it
  Chandra} error circle and reported it
to be a reddened emission-line star. 

%
   \begin{figure}
   \centering
   \resizebox{\columnwidth}{!}{\includegraphics[angle=+90]{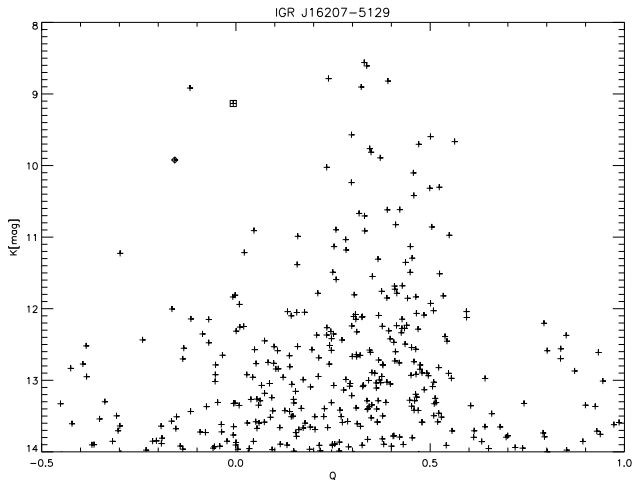}}
   \caption{A plot of IR $Q$ values against $K_{{\rm S}}$ magnitude
   for 2MASS stars within $3\arcmin$ of the position for
   IGR~J16207$-$5129. Candidate early-type stars clearly stand out on
   the top left corner. The object marked with a diamond is
   J16203633$-$5129352 = HD 146803, while the object marked with a
   square is J16204627$-$5130060, the correct counterpart.  The
   brightest candidate, J16202603$-$5129275, seems to be resolvable
   into two objects. Typically, errors in $K$ run from $\sim 0.02$ for the
   brightest objects to $\sim 0.04$ at $K_{{\rm S}}=14$, but can be
   rather larger for blended objects (we did not select candidates
   with $\Delta K>0.05$). Errors in $Q$ vary from $\Delta Q=0.04$ to
   rather high values, with a mean value of $\Delta Q=0.11$ and a
   median $\Delta Q=0.10$.} 
              \label{diagram}
    \end{figure}

This source offers a suitable example for the application of the $Q$
method. Fig.~\ref{diagram} shows a plot of $Q$ against $K_{{\rm S}}$
magnitude for all stars brighter than $K_{{\rm S}}= 14$ within a
$3\arcmin$ radius from the nominal position of IGR~J16207$-$5129. The
four stars fulfilling our criteria 
 stand clearly on the left top corner of
the plot. However, the magnitudes of the object with $Q=-0.30$,  2MASS
J16202935$-$5130457 are flagged as bad. Therefore its $Q$ value is
likely to be meaningless, and we do not consider it a real candidate
to be a Be star. The bright source 2MASS 
J16203633$-$5129352 corresponds to the catalogued emission-line star
\object{HD 146803}, reported by \citet{mas06} to be a late Be
star and so not a likely candidate. This leaves us with
two viable candidates. 2MASS J16202603$-$5129275 has colours and
magnitudes typical of an early-type supergiant. However, close
examination of the images available suggests that the optical source
USNO-B1.0 0385-0554515 is not the same object, as its position is
displaced by $\sim 2\arcsec$ and DENIS identifies two separate objects
here, one coincident with the 2MASS source and one coincident with the
USNO source. This means that the $J$ mag is likely to represent a
blend of both objects and so, again, the $Q$ value is meaningless.

The only candidate left, 2MASS J16204627-5130060 = 
\object{USNO-B1.0 0384-0560875}, was confirmed as the counterpart to
IGR~J16207$-$5129 by a {\it Chandra} localisation shortly before our
run \citep{tomsick,mastel}. USNO-B1.0 quotes $B=19.7$ for this object,
far too faint for 
  a classification spectrum. However, the $I$-band spectrum can be
  used to classify luminous OB stars following the prescriptions of
  \citet{caron} and \citet{c05}.

\begin{figure}
   \centering
\resizebox{\columnwidth}{!}{\includegraphics[bb= 70 20 540 820]{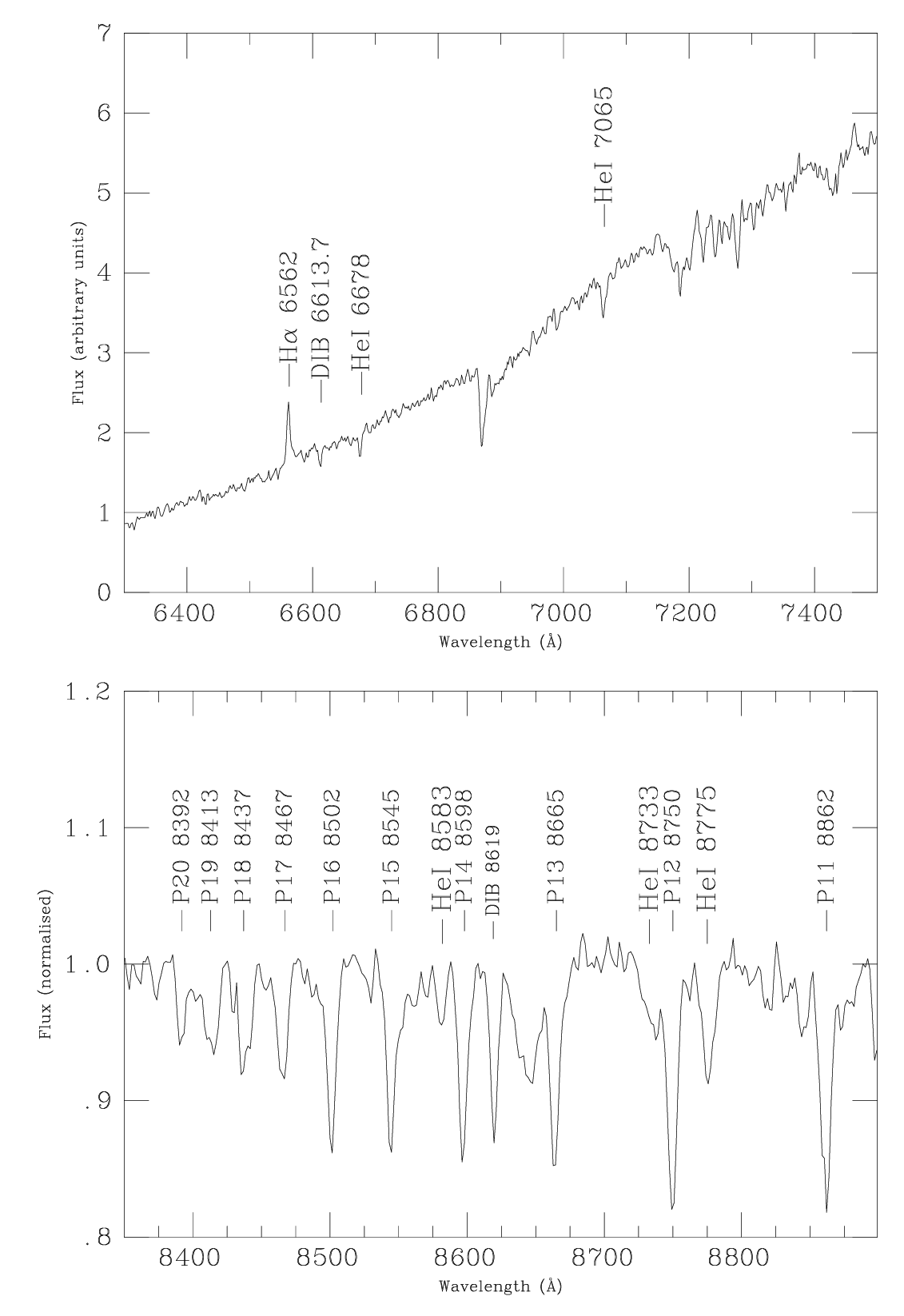}}
   \caption{Spectrum of USNO-B1.0 0384-0560875, the optical
   counterpart to IGR~J16207$-$5129. The top panel shows the red part
   of the spectrum, while the lower panel zooms in the region around
   the Paschen lines. The features in this region clearly identify the
   star as an early B supergiant.}
              \label{sgs}
    \end{figure}

We observed \object{USNO-B1.0 0384-0560875} in the red and far
red. Its spectrum is shown in Fig.~\ref{sgs}. The spectrum shows deep
and well separated Paschen lines, several \ion{He}{i} lines and no
obvious metallic lines. The Paschen lines are well separated and 
resolvable up to at least Pa~20. The \ion{He}{i} lines are resolved
even at this low resolution. These features
clearly identify the object as an OB supergiant. The lack of
\ion{O}{i}~8446\AA\ and other metallic lines makes it earlier than
B3. Moreover, the lack
of \ion{O}{i}~7774\AA\ indicates that it is earlier than B1. A limit
on the hot side is more difficult to set. However,
\ion{He}{ii}~6683\AA\ becomes clearly resolved from
\ion{He}{i}~6678\AA\ around O8.5 and there is no sign of it in our
spectrum. Moreover, the Paschen lines tend to become shallower and
less well defined in O-type supergiants. Therefore we conclude that
this object has a spectral type of B0 with an uncertainty of about one
subtype.

%
%

In view of its optical counterpart and persistent nature,
\object{IGR~J16207$-$5129} is a new SGXB.

\subsection{IGR~J16283$-$4843}

This source has been observed by different X-ray observatories
\citep{beck05}. It presents a hard spectrum, with very high intrinsic
absorption. The X-ray spectrum strongly suggests that it is a HMXB
\citep{beck05}. The {\it SWIFT} error circle encompasses a single 2MASS
source, 2MASS J16281083-4838560. This object is in reality a blend of
images of
at least three stars \citep{steeghs} that coincides with a {\it
  Spitzer} source \citep{beck05}. The mid-IR detection favours the
identification of this source as the counterpart, but its faintness
($K=13.95$) is surprising for the counterpart of a HMXB.

Because of this, we searched the vicinity of this position and found a
bright 2MASS source, $\approx 9\arcsec$ away from the centre of the
error circle, 2MASS J16280983$-$4838590. This object is relatively
bright $K=12.01$ and has $Q=-0.3$, fulfilling our conditions for a
candidate Be star.  However, deep $K$-band images of this area by
\citet{steeghs} show that there are several fainter $K$-band sources
surrounding 2MASS J16280983$-$4838590 and suggest that this source is
actually a blend of two objects of similar magnitudes.

A coincident optical source, USNO-B1.0 0413-0504855 ($R_{1}=14.2$)
could correspond to 2MASS J16280983$-$4838590, but it looks far too
bright in the optical to be the same as an object with $(J-K_{\rm
  S})=1.3$. In any event, we observed it and it turns out to be a
K-type star. Obviously, a deep
IR investigation of this error circle is necessary.

\subsection{IGR~J16320$-$4751}

This {\it INTEGRAL} source, also known as  AX J1631.9-4752, is a
  $P_{{\rm s}}=1300\,{\rm s}$ pulsar \citep{rod06,lut05a}.  Monitoring
  with the {\it SWIFT}/BAT instrument has
resulted in the discovery of an 8.96-d periodicity in its light
  curve \citep{cor05}. The X-ray spectrum of \object{IGR~J16320$-$4751} 
  suggests it is a HMXB, while the periodicities detected are typical
  of SGXBs.  There is one source inside the  {\it
  XMM-Newton} error circle, 2MASS J16320175$-$4752289, and a second
  one just outside. 2MASS J16320175$-$4752289, is rather faint. 2MASS gives
  $K=10.99$, $H=13.03$ and only an upper limit in $J$. Therefore we
  could not calculate its $Q$ value. The other source, 2MASS
  J16320215$-$4752322 is visible in the optical as USNO-B1.0
  0421-0625270 (with $R_{2}=15.4$). This object, with  
  $K=10.8$ and $Q=0.18$, is marginally consistent with being a reddened OB
  supergiant. However, our spectrum shows that  2MASS
  J16320215$-$4752322 is a K-type giant. 
This suggests that 2MASS J16320175$-$4752289 is the actual counterpart
to this HMXB, as seems to be confirmed by the report by
  \citet{wal06} that it is an early-type star. If it is an OB
  supergiant, it has to be extremely reddened.

\subsection{AX J1700.2$-$4220}
An {\it INTEGRAL} source coincident with this faint {\it ASCA} source
is listed in the 1st {\it INTEGRAL} catalogue
\citep{bird04}. \citet{mastel} mention that the catalogued Be star
\object{HD~153295} is within the {\it INTEGRAL} 
error circle. However, it is outside the smaller {\it ASCA} error
circle. Hence we looked for other candidates within the {\it ASCA}
error box.  Analysis of the 2MASS stars within the error circle gives two good
candidates to be emission-line early-type stars within the {\it ASCA}
source and one further candidate to be a supergiant just outside
it (see Table~\ref{tab:no-one}).

   \begin{table}
      \caption[]{Candidate blue stars in the field of AX
      J1700.2$-$4220. Note that the third candidate is formally
      outside the {\it ASCA} error circle. 
}
         \label{tab:no-one}
\begin{tabular}{c c c c c}        
\hline\hline   
2MASS & USNO-B1.0& $Q$ & $K$ & $R_{2}$\\
            \hline
J17001950$-$4219410&0476-0587824&$-$0.15 & 11.65 &13.7\\ 
J17001531$-$4219395&0476-0587760&$-$0.08 &11.30&14.3 \\
J17001663$-$4219078&0476-0587779&+0.02& 10.58&13.4\\
            \hline
        \end{tabular}
   \end{table}
%
%

The low resolution spectrum of 2MASS J17001531$-$4219395 shows both a
 strong \ion{Ca}{ii} triplet and  
deep, well defined Paschen lines, suggesting that this is a late A/early
F star of at least moderate luminosity. We obtained a blue spectrum,
which, though rather noisy, suggests that the spectral type is not far
from F2\,III. 
2MASS J17001950$-$4219410 shows a broad H$\alpha$ absorption line and
some weak metallic 
lines, indicative of a late A-type star, likely of moderate
 luminosity. Though both stars are relatively early, none is a
 believable candidate. From the richness of metals in its spectrum and
 the strength of the 
metallic blend around 6495\AA, 2MASS J17001663$-$4219078 is a G-type
 star. The reddening in this direction is perhaps anomalous, as all
 three stars appear to show $Q$ parameters that are much more
 negative than the values that
 correspond to their spectral types.

The lack of convincing candidates within the {\it ASCA} error circle
opens up the possibility that \object{HD~153295} is the actual
counterpart. As expected for a bright Be star, HD~153295 fulfills our
conditions with $K=6.73$ and $Q=-0.26$. We obtained intermediate
resolution spectra of this 
star. The red spectrum shows several weak metallic 
lines and a prominent unresolved H$\alpha$ emission line with EW =
$-68\pm2$\AA. This is quite a high value for a Be star. The blue
spectrum (see Fig~\ref{bes}) confirms the strong Be
characteristics. The spectral type is difficult to assign because of
the presence of many weak emission lines, but the observed absorption
lines place it not very far away of B0.5\,IVe. \citet{gar77} give
B1?\,III, in rather good agreement. There are several $UBV$
photometric measurements for this star, showing very little
variability. Taking, for example, \citet{schild}, we have $V=9.04$,
$(B-V)=0.55$. The strong Be characteristics of \object{HD~153295}
imply an important contribution
of the circumstellar disk to both $E(B-V)$ and $V$,
likely on the order of $E(B-V)_{{\rm disk}}\approx0.15$ and
$\Delta V_{{\rm disk}}\approx0.5$ \citep[cf.][]{dachs88}.

Assuming standard colours and
magnitudes for  \object{HD~153295}, $(B-V)_{0}=-0.24$ \citep{weg94}
and $M_{V}=-4.5$ 
\citep{hme84},  and a standard reddening law, we derive
$d\approx1.7\:$kpc (a similar value is obtained from the 2MASS
magnitudes). This is actually a lower limit to the distance. If we
assume the likely contribution from the circumstellar disk given
above, we obtain $d\approx2.6\:$kpc This distance is comparable to
estimates around $2\:$kpc for the nearby cluster NGC~6231
\citep[e.g.,][and references therein]{rab97}, believed 
to be the core of the Sco OB1 association. If we were to adopt a
luminosity class V for HD~153295, the distance would be fully
compatible with membership in Sco OB1.

If HD~153295 is indeed an
outlying member of Sco OB1, it may be too young to be
a Be/X-ray binary. NGC~6231 is estimated to be very young
($\sim4\:$Myr), while typical evolutionary timescales to form Be/X-ray
binaries are $\sim 10\:$Myr. \citet{the94} have discussed the possible
association of HD~153295 with the IRAS source 16569$-$4213. If the
association is true, HD~153295 would be a Herbig Be star and certainly
not an X-ray binary, but this would imply an age much younger
than $\sim4\:$Myr (contraction times for a B0 star are
$\sim10^{5}$~yr; cf. \citealt{ps99})\footnote{Also, there is another
  emission line star 
  in the area that could be associated with the IRAS source, GSC
  07877-00412.}. Clearly, a better position for AX J1700.2$-$4220 is
necessary before the association with HD~153295 can be considered with
any confidence.
 
\subsection{IGR~J17091$-$3624}

First seen by {\it INTEGRAL} in April 2003 \citep{kuul},
\object{IGR~J17091$-$3624} 
was later found in archival observations taken by several other
satellites \citep{capi06}. The source is very variable in flux and
spectral index, and shows little absorption. Its spectrum
can be very hard or relatively soft, suggesting it is a black hole
\citep{lut05b,capi06}. \citet{rup03} found a variable radio source within the
error circle with the VLA. Though these findings suggest that
IGR~J17091$-$3624 is a low-mass microquasar, we were intrigued to find
that the only candidate blue star in the error circle, 2MASS
J17090199-3623260, was within the VLA source error circle. 

Later, \citet{pan06} observed a radio source at a position coincident
with that reported by \citet{rup03}. According to \citet{pan06}, the
only possible counterpart to these sources is 2MASS
J17090199-3623260. This object is also detected in the optical as
USNO-B1.0 0536-0466988. It has $K=11.7$, $Q=0.0$ and $R_{1}=14.46$
It should be noted that both the $J$ and $K$ magnitudes for 2MASS
J17090199-3623260 are tagged as upper limits. However, they are
fully compatible within the errors with the corresponding DENIS
magnitudes. 

%
   \begin{figure}
   \centering
\resizebox{\columnwidth}{!}{\includegraphics[]{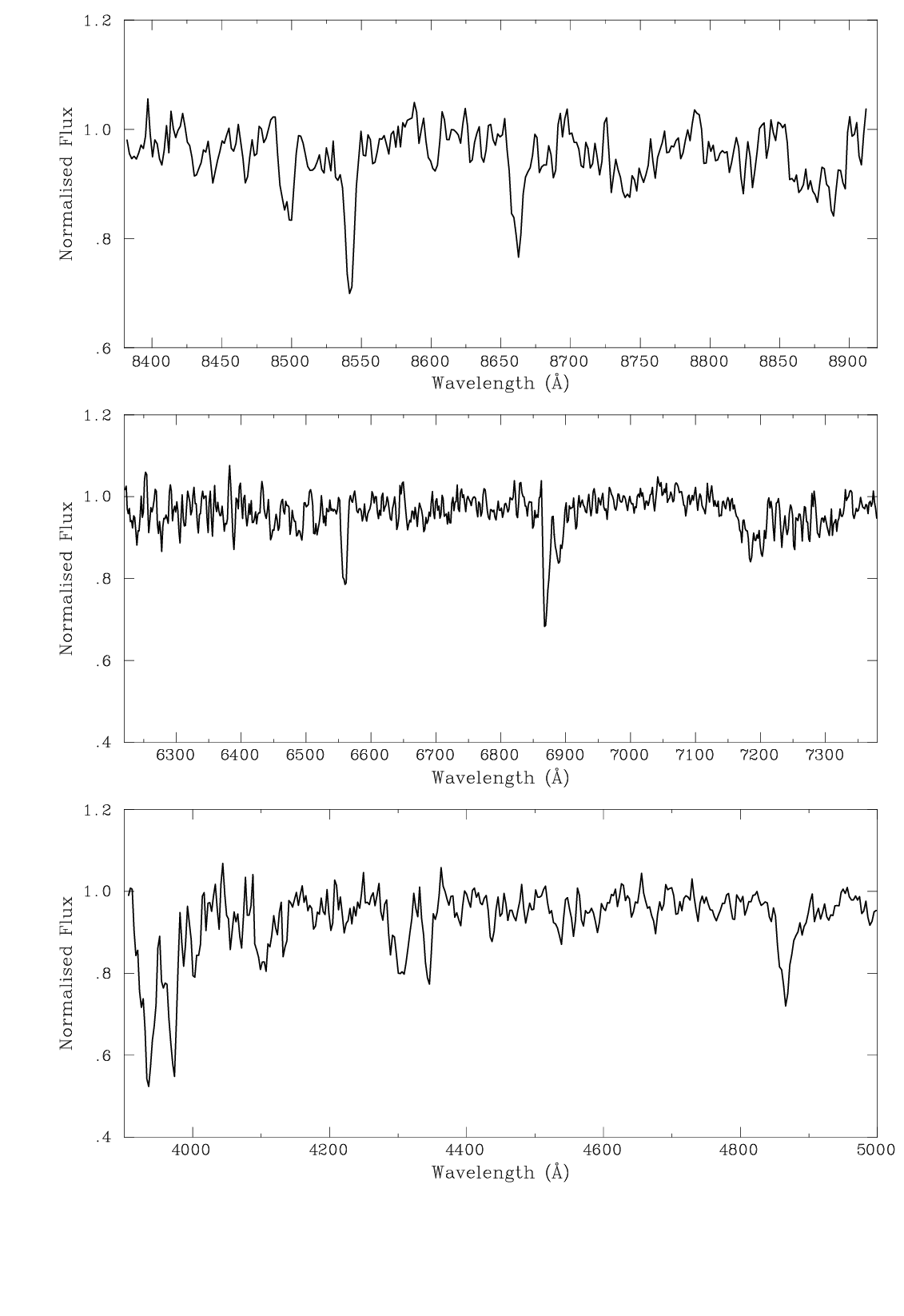}}
   \caption{Several regions of the spectrum of USNO-B1.0 0536-0466988, likely associated
   with IGR~J17091$-$3624. While the blue and $I$-band spectra
   indicate a late F type, the red spectrum looks more typical of a
   rather later spectral type, late G or early K.}
              \label{mq}
    \end{figure}
%
	
The spectrum of USNO-B1.0 0536-0466988 is displayed in
Fig~\ref{mq}. The $I$-band spectrum shows prominent narrow \ion{Ca}{ii}
triplet lines and weak Paschen lines, suggesting a mid or late F-type
star of moderate luminosity.  The low resolution spectrum taken with
the NOT covers the 
blue range. Based on the features present, we estimate a spectral type
around F8\,V. However, the spectrum around H$\alpha$ resembles a
rather later star. In any case, this object is not an early
type star and does not show any sign of emission lines. 

In view of the almost secure association with the radio source, we may
speculate that IGR~17091$-$3264 could be a low-mass transient
microquasar observed in quiescence. According to
\citet{capi06}, the source has not been active since 2004.
Perhaps the $JHK$ observations
were made at a time when the source was active and the colours
dominated by an accretion disk, mimicking the flat IR spectrum of an
early type star. We must note, however,
that there there is a second faint 2MASS source partially blended with
our target, 2MASS J17090219-3623292, which has $H=13.84$ and
uncertain $J$ and $K$ magnitudes.

If the 2MASS magnitudes are intrinsic to the star, we can estimate
its distance, assuming $M_{V} = +3.5$ for an F8\,V star \citep{mh82}
and $(V-K)_{0}=1.12$ \citep{koo85}. By using $E(J-K_{{\rm S}}) = 1.3
A_{K_{{\rm S}}}$ (assuming the extinction law of \citealt{rl85}), we obtain
$A_{K_{{\rm S}}}\approx0.2$ and hence $d\sim 800\:$pc. This is
certainly much closer than the Galactic Bulge, where most low mass
  X-ray binaries are known to 
  reside and this source was suspected to lie
  \citep[e.g.][]{capi06}. Even if USNO-B1.0 
0536-0466988 is the actual counterpart, and the 
2MASS magnitudes were obtained during a bright state and they come
from an accretion disk, the source is very unlikely to be as far as the
Galactic Bulge.

\subsection{AX~J1820.5$-$1434}

This source was discovered by {\it ASCA} in 1997 \citep{kin98}, as a
moderately faint X-ray pulsar ($P_{{\rm s}}=152.3\:{{\rm s}}$). Its
X-ray spectrum is highly absorbed, while its
transient nature suggests a Be/X-ray binary. {\it INTEGRAL} has
detected this source up to 70~keV \citep{fil05}.

Analysis of the 2MASS data for the error circle reveals a single
candidate to being an early type star, 2MASS J18203114$-$1434193. This
object has $K=10.98$ and $Q=0.06$. The corresponding optical source is
USNO-B1.0 0754-0489829, with $R_{2}=14.7$ and $B_{2}=16.5$.

Our spectrum of this source has a very low signal to noise
ratio. H$\alpha$ is seen as a broad absorption feature (with perhaps
some emission infilling, though this could be due to the noise) and there are
no indications of metallic lines. 2MASS J18203114$-$1434193 is hence
an early type star, but its 
faintness prevented us from taking a blue spectrum from SAAO. 
A low resolution
spectrum was taken later from the NOT. The spectrum is of moderate
signal to noise and low resolution, but the star is certainly a B-type
star. The object is not a late B star, as several \ion{He}{i} lines
are clearly present, but a better
spectrum will be necessary to determine an accurate spectral type. 

In the NOT spectrum, H$\alpha$ is seen in absorption. However, many
Be/X-ray binaries are known to spend part of the time in
absorption-line phases. Therefore, if 2MASS J18203114$-$1434193 is an
early B-type star, it should be considered as a possible counterpart
for AX J1820.5$-$1434.

\subsection{IGR~J18406$-$0539}

The existence of IGR~J18406$-$0539 is dubious. It is located only $3\farcm5$
from the SFXT AX~J1841.0$-$0535 = IGR~J18410-0535. So, though formally
the error circles for both sources do not overlap (AX~J1841.0$-$0535
has a {\it Chandra} localisation), there is a non-negligible chance that
they are the same source \citep{mas4}. This is a very rich field, but
\citet{mas4} note the presence of the catalogued emission-line star
\object{SS~406} on the edge of the error circle. From a low resolution
spectrum, they deduce that it is a mid Be star, which, they argue,
could be associated with IGR~J18406$-$0539.

%
   \begin{figure*}
   \centering
   \resizebox{\textwidth}{!}{\includegraphics[bb= 80 80 440 760, angle=-90]{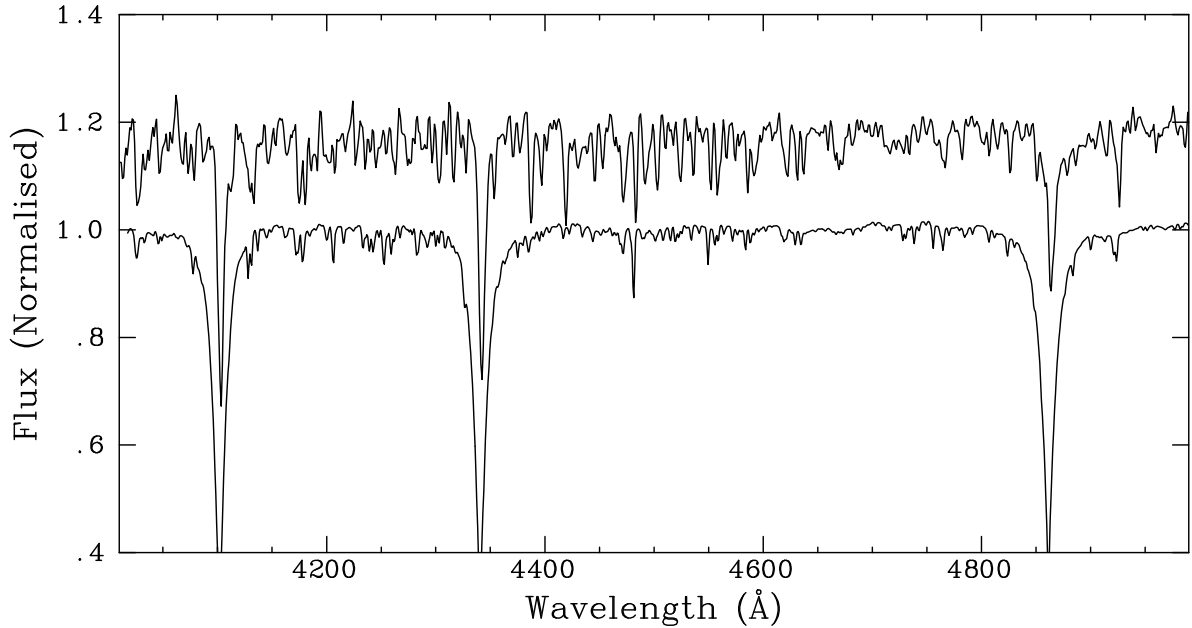}}
   \caption{The blue spectrum of SS~406 (top) compared to the Mn-Hg
   B8p star HD~221507  (a binned version of a UVES spectrum from the
   UVES Paranal Observatory Project; \citealt{bag03}). SS~406 has much
   narrower Balmer lines, but the 
   species present in both spectra mostly match. The presence of
   stronger \ion{He}{i} lines suggests that SS~406 is slightly earlier.}
              \label{weirdb}
    \end{figure*}
%

 The observed spectrum of SS~406 (see Fig.~\ref{weirdb}) is
   strange. 
The H$\alpha$ spectrum, obtained on the night of May 5th, displays a
moderately strong H$\alpha$ emission line, shallow \ion{He}{i}~6678
and 7065\AA\ lines and some sharp metallic lines. The blue
spectrum, taken 3 nights later, displays very narrow Balmer lines, weak sharp
\ion{He}{i} lines and a wealth of metallic lines. The observed
metallic lines do not match a typical F-G spectrum, but rather suggest
that SS~406 is a chemically peculiar B-type star. 
Comparison of the observed metallic spectrum with those of several CP
stars suggests that SS~406 is a Mn-Hg star, while the narrowness of
its Balmer lines would point to a moderate luminosity.

Chemically peculiar stars are known to be slow rotators, while Be
stars are fast rotators. Moreover, the moderately
strong \ion{He}{i} lines seen in the red spectrum seem incompatible with a
late-B CP star. We can think of two explanations for this combination
of spectral features. Perhaps SS~406 is a binary
composed of a mid Be star and a late-B CP giant. Alternatively, a CP
star may show a rotationally modulated spectrum, with \ion{He}{i} and
metallic lines changing in appearence and intensity. Whatever the case, SS~406
does not look like a promising counterpart to IGR~J18406$-$0539, if
  this is indeed a real source. Even if there is a Be star in
  the system, it is unlikely to be sufficiently early to fall in the
  range observed for Be/X-ray binaries. Moreover, the evolutionary
  history of such a system is difficult to conceive.

The $Q$ method reveals a wealth of possible
early-type counterparts within the error circle of
  IGR~J18406$-$0539, though many of them have magnitudes flagged 
as bad. A correct identification of this source must
wait for a better localisation. We note that no X-ray source is
visible in this area in the {\it Chandra} pointing that led to the
localisation of AX~J1841.0$-$0535.

\subsection{Other fields}
Our method did not find any suitable candidate in the field of the
SFXT IGR J16479$-$4514.  We did not find any reliable candidates in
the field of the SGXB IGR~J16418$-$4532. In this field, few 2MASS
stars appear unblended. We observed two very marginal candidates (both
with $Q=0.15-0.20$) whose 2MASS magnitudes were flagged as bad and both
turned out to be late F-type stars. Finally, no candidate blue stars
are found in the field of the hard transient IGR~J11321$-$5311.

\begin{table*}
\caption{Summary of results. The top panel contains data for the
  proposed counterparts to five sources detected with {\it INTEGRAL}
  which had been identified before this work. References are
  \citet{fc04}, \citet{neg05}, \citet{neg}, \citet{pell} and
  \citet{ns06} respectively. The middle panel summarises
  the results presented here. The bottom panel presents data for the
  counterparts to four other {\it INTEGRAL} sources proposed as HMXBs
  by \citet{wal06}.}             
\label{tab:final}      
\centering          
\begin{tabular}{l c c c c c c c c c}     
\hline\hline       
X-ray source& 2MASS& USNO B1.0& $Q$ & $K_{{\rm S}}$& $B_{1}$ & $R_{1}$&
Counterpart& Spectral&Comments\\
&source& source& &&&&Correct&Type&\\
\hline  
IGR J16318$-$4848& J16314831$-$4849005&...& $-0.11$& 7.19&...&...&yes&B[e]&obscured\\                 
IGR J16465$-$4507& J16463526$-$4507045& 0448$-$0520455& $+0.06$&
9.84&15.2 ($B_{2}$)&12.7&yes&B0.5\,I&SFXT?\\ 
XTE J1739$-$302& J17391155$-$3020380 &0596$-$0585865& $+0.07$& 7.43&17.3&13.2&yes&O8.5\,Iab(f)& SFXT\\ 
IGR J17544$-$2619& J17542527$-$2619526& 0636$-$0620933& $-0.02$& 8.02&14.5&11.3&yes&O9\,Ib&SFXT\\
SAX J1818.6$-$1703& J18183790$-$1702479& 0729$-$0750578& $+0.09$& 7.85&...&16.8&likely&O9-B1\,I&SFXT\\
\hline
IGR J11305$-$6256& J11310691$-$6256489& 0270$-$0309619& $-0.12$& 8.01& 8.1& 8.2&yes&B0.5\,IIIe&HD~100199\\
IGR J16207$-$5129& J16204627$-$5130060& 0384$-$0560875& $-0.01$& 9.13&...&15.2&yes&B0\,I&SGXB\\
IGR J16283$-$4843& J16280983$-$4838590& 0413$-$0504855& $-0.30$&
12.01&...&14.0&no&K&...\\
IGR J16320$-$4751&J16320175$-$4752289&...&...&10.99&...&...&likely&...&...\\
&J16320215$-$4752322&0421-0625270&$+0.18$&10.82&$17.3(B_{2})$&14.6&no& K& ...\\
AX J1700.2$-$4220& J17002524$-$4219003& 0476$-$0587932& $-0.26$& 6.73& 9.5& 8.8&perhaps&B0.5\,IVe&HD~153295\\
IGR J17091$-$3624& J17090199$-$3623260& 0536$-$0466988& $+0.00$& 11.7&...& 14.46&likely&F8\,V&Radio source\\
AX J1820.5$-$1434& J18203114$-$1434193& 0754$-$0489829& $+0.06$&
10.98& $16.5(B_{2})$&$14.7(R_{2})$&perhaps&B&not Be\\
IGR J18406$-$0539& J18404689$-$0540502& 0843$-$0409384& $-0.05$& 8.613& 12.7&11.4&unlikely&Mn-Hg&SS~406\\
\hline
IGR J16393$-$4641& J16390535$-$4642137&...& $+0.41$&12.78&...&...&candidate&...&...\\
IGR J16418$-$4532& J16415078$-$4532253&...& $+0.16$&11.48&...&...&candidate&...&...\\
IGR J16479$-$4514& J16480656$-$4512068& 0447$-$0531332& $+0.38$& 9.80&...&18.36&candidate&...&...\\
IGR J17252$-$3616& J17251139$-$3616575&...& $+0.47$& 10.67&...&...&candidate&...&...\\
\hline                  
\end{tabular}
\end{table*}

\section{Conclusions}
 
   \begin{enumerate}
     \item We have used existing photometric catalogues to search for
     early type stars within the error circles of X-ray sources
     believed to be high mass X-ray binaries. Our main discriminant
     has been the infrared $Q$ parameter, while optical magnitudes
     from USNO B1.0 have been used in an attempt to rule out
     foreground A or F-type 
     stars. The method has proved efficient at finding reddened
     OB stars, resulting in the detection of the likely counterpart to
     SAX~J1818.6$-$1703 \citep{ns06} and the selection of the counterpart to
     IGR~J16207$-$5129, later identified through a {\it Chandra}
     pointing. However, it does not allow the discrimination of reddened
     A or F-type stars. We have found that many bright candidates
     selected with $Q\sim 0.1$ 
     turn out to be reddened A or F giants. Most sources with $Q=0.15-0.20$
     are reddened late F or G stars. Surprisingly, in a few cases, the
     candidates turn out to be late-type stars. The most likely
     explanation for these misidentifications is that the 2MASS
     sources must be 
     unresolved blends and therefore the $Q$ values calculated lack
     astrophysical meaning.
     \item Recently, \citet{wal06} have reported on {\it XMM-Newton}
     localisations of 10 {\it INTEGRAL} sources, nine of which they
     believe to be SGXBs. It is worth noting that, from those 9
     sources, the three objects with confirmed early-type luminous
     counterparts, IGR J16318$-$4848, IGR J16465$-$4507 and IGR
     J17544$-$2619, would have been selected as candidate OB
     supergiants by our method, but the proposed candidates for IGR
     J16393$-$4641, IGR J16418$-$4532, IGR J16479$-$4514 and IGR
     J17252$-$3616 would not (see Table~\ref{tab:final}).
     The proposed candidates for IGR~J16320$-$4751 and IGR
     J18027$-$2016 lack reliable $J$ magnitudes and so $Q$ values could not
     be calculated.
     
While, for example, the proposed counterpart to IGR J16465$-$4507
     fails the $Q$ value criterion (perhaps indicating contamination
     of the 2MASS magnitudes by other nearby sources), the proposed
     counterparts to several sources fail the magnitude limit. In the
     cases of  IGR~J16283$-$4843 and
     IGR~J16320$-$4751, the proposed counterparts are very faint in
     $K$. Taking into
     account the huge intrinsic luminosity of an OB supergiant, these
     counterparts must be very heavily absorbed not only in the
     optical, but also in the infrared. If this absorption was mostly
     intrinsic to the sources (i.e., a very thick wind or dusty
     envelope surrounding the system), we would expect the systems to
     be bright mid-IR sources. This is certainly the case in IGR
     J16318$-$4848 \citep{fc04} and could also happen for
     IGR~J16283$-$4843, perhaps associated with a mid IR
     source. However, in other cases, like IGR~J16320$-$4751, there is
     no such mid IR emission and it seems very unlikely that the
     absorption occurs in the system itself. The possibility that
     these objects are HMXBs with main-sequence primaries should be
     considered. \citep[cf.][]{pfahl,ribo}.  
     \item Our method reveals a B-type star inside the error circle
     for the X-ray pulsar AX~J1820.5$-$1434, but our spectra are not
     good enough to 
     decide if this is an early B-type star (and hence a potential
     candidate). H$\alpha$ seems to be in absorption.
     \item For HD~100199, identified with IGR~J11305$-$6256, we
     confirm the spectral type B0\,IIIe.
     \item We fail to find any suitable counterpart within the {\it
     ASCA} error circle for AX J1700.2$-$4220. For the suggested
     counterpart, HD~153295, which is inside the {\it INTEGRAL} error
     circle, but not the {\it ASCA} circle, we find a spectral type
     B0.5\,IVe and a distance compatible with membership in the Sco
     OB1 association. 
     \item In the error circle for the likely black hole transient
     IGR~J17091$-$3624, a variable radio source has been detected. The
     star likely associated with this radio source is $\sim$F8\,V and
     must be much closer than the Galactic Bulge.
     \item We rule out the association of SS~406 with
     IGR~J18406$-$0539, as this object is a late B chemically peculiar
     star, and perhaps a binary consisting of two unevolved
     stars. 
   \end{enumerate}

\begin{acknowledgements}
      
We would like to thank Andrew~J.~Norton for helpful comments on the
manuscript and the referee, Doug Gies, for many valuable suggestions
that improved notably the quality of the paper.  
IN is a researcher of the programme {\em Ram\'on y Cajal}, funded by
the Spanish Ministerio de Educaci\'on y
Ciencia and the University of Alicante, with partial 
support from the Generalitat Valenciana and the European Regional
Development Fund (ERDF/FEDER).
This research is partially supported by the MEC under
grant AYA2005-00095. During part of this work, IN was a visiting
fellow at the Open University, whose kind hospitality is warmly
acknowledged. This visit was funded by the MEC under grant
PR2006-0310.   

This research has made use of the Simbad data base, operated at CDS,
Strasbourg (France). This publication makes use of data products from
the Two Micron All 
Sky Survey, which is a joint project of the University of
Massachusetts and the Infrared Processing and Analysis
Center/California Institute of Technology, funded by the National
Aeronautics and Space Administration and the National Science
Foundation. It also makes use of DENIS, the result of a joint effort
involving human and financial 
    contributions of several Institutes mostly located in Europe. It has
    been supported financially mainly by the French Institut National des
    Sciences de l'Univers, CNRS, and French Education Ministry, the
    European Southern Observatory, the State of Baden-Wuerttemberg, and
    the European Commission under networks of the SCIENCE and Human
    Capital and Mobility programs, the Landessternwarte, Heidelberg and
    Institut d'Astrophysique de Paris. This paper makes use of data
    from the UVES Paranal Observatory Project (ESO 266.D-5655).

Part of the data presented here have been taken using ALFOSC, which is 
owned by the Instituto de Astrof\'{\i}sica de Andaluc\'{\i}a (IAA) and
operated at the Nordic Optical Telescope under agreement
between IAA and the NBIfAFG of the Astronomical Observatory of
Copenhagen. 
\end{acknowledgements}

\end{document}